\documentclass[preprint]{jpconf}
\usepackage{amssymb}
\usepackage{amsmath}
\usepackage{citesort}

\begin{document}

\title{Fermions embedded in a scalar-vector kink-like smooth potential\footnote{J. Phys.: Conf. Series 630 (2015) 012029}}
\author{W M Castilho and A S de Castro}
\address{UNESP, Campus de Guaratinguet\'{a}, Departamento de F\'{\i}sica e
Qu\'{\i}mica, \protect{12516-410} Guaratinguet\'{a}, SP,  Brazil}

\begin{abstract}
The behaviour of massive fermions is analyzed with scalar and vector
potentials. A continuous chiral-conjugation transformation decouples the
equation for the upper component of the Dirac spinor provided the vector
coupling does not exceed the scalar coupling. It is shown that a
Sturm-Liouville perspective is convenient for studying scattering as well as
bound states. One possible isolated solution (excluded from the
Sturm-Liouville problem) corresponding to a bound state might also come into
sight. For potentials with kink-like profiles, beyond the intrinsically
relativistic isolated bound-state solution corresponding to the zero-mode
solution of the massive Jackiw-Rebbi model in the case of no vector
coupling, a finite set of bound-state solutions might appear as poles of the
transmission amplitude in a strong coupling regime. It is also shown that
the possible isolated bound solution disappears asymptotically as the
magnitude of the scalar and vector coupling becomes the same. Furthermore,
we show that due to the sizeable mass gain from the scalar background the
high localization of the fermion in an extreme relativistic regime is
conformable to comply with the Heisenberg uncertainty principle.
\end{abstract}

\section{Introduction}

The Dirac Hamiltonian with a mixing of scalar potential and time component
of vector potential in a four-dimensional space-time is invariant under an
SU(2) algebra when the difference between the potentials, or their sum, is a
constant \cite{Bell1975151}. The near realization of these symmetries may
explain degeneracies in some heavy meson spectra (spin symmetry) \cite%
{PhysRevLett.86.204,Ginocchio2005165} or in single-particle energy levels in
nuclei (pseudospin symmetry) \cite%
{Ginocchio2005165,PhysRevLett.78.436,Ginocchio19981,PhysRevC.58.R45,PhysRevC.58.R628,PhysRevC.58.R3065,Ginocchio1999231,PhysRevC.62.054309,PhysRevLett.86.5015,Marcos200130,PhysRevLett.87.072502,PhysRevC.65.034307,PhysRevC.66.064312,0256-307X-20-3-312,PhysRevLett.91.262501,PhysRevC.67.044318,PhysRevC.69.024319,PhysRevLett.92.202501,Guo200590,PhysRevC.71.034313,PhysRevC.72.054319,Guo2005411,Berkdemir200632,Xu2006161,he,jolos,0256-307X-26-12-122102,liang,PhysRevC.81.064324,1674-1137-34-9-061,PhysRevC.83.041301,0256-307X-28-9-092101,PhysRevLett.109.072501,PhysRevA.86.032122,PhysRevC.87.031301,1742-6596-490-1-012069}%
. When these symmetries are realized, the energy spectrum does not depend on
the spinorial structure, being identical to the spectrum of a spinless
particle \cite{PhysRevC.75.047303}. Despite the absence of spin effects in
1+1 dimensions, many attributes of the spin and pseudospin symmetries in
four dimensions are preserved.

In a pioneering work, Jackiw and Rebbi \cite{PhysRevD.13.3398} have shown
that massless fermions coupled to scalar fields with kink-like profiles in
1+1 dimensions develops quantum states with fractional fermion number due to
the zero-mode solution. This phenomenon has been seen in certain polymers
such as polyacetylene \cite%
{PhysRevLett.42.1698,goldstone1981fractional,PhysRevLett.50.439,niemi1986fermion}%
. Recently the complete set of solutions for the kink-like scalar potential
behaving like $\mathrm{\tanh }\,x\!/\!\lambda $ has been considered for
massless fermions in Ref. \cite{PhysRevD.89.025002}, and for massive
fermions in Ref. \cite{Charmchi2014256}. The complete set of solutions for
massive fermions under the influence of a kink-like scalar potential added
by the time component of a vector potential with the same functional form
was considered in Refs. \cite{Castilho20141} and \cite{Castilho2014164}, in
Ref. \cite{Castilho20141} for the background field behaving like $\mathrm{sgn%
}\,x$, and in Ref. \cite{Castilho2014164} for the background behaving like $%
\mathrm{\tanh }\,x\!/\!\lambda $.

In Refs. \cite{Castilho20141} and \cite{Castilho2014164}, it has been shown
that the Dirac equation with a scalar potential plus a time component of
vector potential of the same functional form is manageable if the vector
coupling does not exceed the scalar coupling, and that the bound states for
mixed scalar-vector potentials with kink-like profiles are intrinsically
relativistic solutions. Furthermore, it has been shown that the fermion can
be confined in a highly localized region of space under a very strong field
without any chance for spontaneous pair production related to Klein's
paradox.

Here we shall outline the scalar-vector mixing framework developed in Refs.
\cite{Castilho20141}-\cite{Castilho2014164} with the Sturm-Liouville
perspective plus its isolated solution. We show that the isolated solution
disappears asymptotically as one approaches the conditions for the
realization of the so-called spin and pseudospin symmetries in four
dimensions. After a general consideration of the Sturm-Liouville problem for
an arbitrary kink-like potential, we concentrate our attention on the
isolated solution which corresponds to the zero-mode solution of the
Jackiw-Rebbi model in the case of a pure scalar coupling. It is shown that
such an isolated solution is an intrinsically relativistic solution even if
the fermion is massive. Then we use the smooth step potential $\mathrm{\tanh
}\,x\!/\!\lambda $ to show in detail how the additional mass acquired by the
fermion from the scalar background can acquiesce highly localized states
without violating the Heisenberg uncertainty principle.

\section{Mixed scalar-vector interactions}

Consider the Lagrangian density for a massive fermion%
\begin{equation}
L=\bar{\Psi}\left( i\hbar c\gamma ^{\mu }\partial _{\mu }-Imc^{2}-V\right)
\Psi  \label{d1}
\end{equation}%
where $\hbar $ is the constant of Planck, $c$ is the velocity of light, $I$
is the unit matrix, $m$ is the mass of the free fermion and the square
matrices $\gamma ^{\mu }$ satisfy the algebra $\{\gamma ^{\mu },\gamma ^{\nu
}\}=2Ig^{\mu \nu }$. The spinor adjoint to $\Psi $ is defined by $\bar{\Psi}%
=\Psi ^{\dagger }\gamma ^{0}$. In 1+1 dimensions $\Psi $ is a 2$\times $1
matrix and the metric tensor is $g^{\mu \nu }=$ diag$\left( 1,-1\right) $.
For vector and scalar interactions the matrix potential is written as%
\begin{equation}
V=\gamma ^{\mu }A_{\mu }+IV_{s}  \label{V}
\end{equation}
We say that $A_{\mu }$ and $V_{s}$ are the vector and scalar potentials,
respectively, because the bilinear forms $\bar{\Psi}\gamma ^{\mu }\Psi $ and
$\bar{\Psi}I\Psi $ behave like vector and scalar quantities under a Lorentz
transformation, respectively. Eq. (\ref{d1}) leads to the Hamiltonian form
for the Dirac equation%
\begin{equation}
i\hbar \frac{\partial \Psi }{\partial t}=H\Psi
\end{equation}%
with the Hamiltonian given as%
\begin{equation}
H=\gamma ^{5}c\left( p_{1}+\frac{A_{1}}{c}\right) +IA_{0}+\gamma ^{0}\left(
mc^{2}+V_{s}\right)
\end{equation}%
where $\gamma ^{5}=\gamma ^{0}\gamma ^{1}$. Requiring $\left( \gamma ^{\mu
}\right) ^{\dag }=\gamma ^{0}\gamma ^{\mu }\gamma ^{0}$, one finds the
continuity equation $\partial _{\mu }J^{\mu }=0$, where the conserved
current is $J^{\mu }=c\bar{\Psi}\gamma ^{\mu }\Psi $. The positive-definite
function $J^{0}/c=|\Psi |^{2}$ is interpreted as a position probability
density and its norm is a constant of motion. This interpretation is
completely satisfactory for single-particle states \cite{0387548831}. The
space component of the vector potential can be gauged away by defining a new
spinor just differing from the old by a phase factor so that we can consider
$A_{1}=0$ without loss of generality.

Assuming that the potentials are time independent, one can write $\Psi
\left( x,t\right) =\psi \left( x\right) \exp \left( -iEt/\hbar \right) $ in
such a way that the time-independent Dirac equation becomes $H\psi =E\psi $.
Meanwhile $J^{\mu }=c\overline{\psi }\gamma ^{\mu }\psi $ is time
independent and $J^{1}$ is uniform.

From now on, we use an explicit representation for the 2$\times $2 matrices $%
\gamma $ as%
\begin{equation}
\gamma ^{0}=\sigma _{3},\quad \gamma ^{1}=i\sigma _{2}
\end{equation}
in such a way that $\gamma ^{5}=\sigma _{1}$. Here, $\sigma _{1}$, $\sigma
_{2}$ and $\sigma _{3}$ stand for the Pauli matrices.

\subsection{Nonrelativistic limit}

The Lorentz nature of the potentials does no matter in a weak-coupling
regime. Indeed, fermions (antifermions) are subject to the effective
potential $V_{s}+A_{0}$ ($V_{s}-A_{0}$) with energy $E\approx +mc^{2}$ ($%
-mc^{2}$) so that a mixed potential with $A_{0}=-V_{s}$ ($A_{0}=+V_{s}$) is
associated with free fermions (antifermions) in a nonrelativistic regime
\cite{Castilho20141}-\cite{Castilho2014164}. The changes of signs of $A_{0}$
and $E$ as well as the invariance of the sign of $V_{s}$ when one exchanges
the roles of fermions and antifermions are justified by the
charge-conjugation transformation.

\subsection{Extreme relativistic limit}

For potentials localized in the range $\lambda $, quantum effects appear
when $\lambda $ is comparable to the Compton wavelength $\lambda _{C}=\hbar
/mc$, and relativistic quantum effects are expected when $\lambda $ is of
the same order or smaller than the Compton wavelength. The fermion is under
an extreme relativistic regime if $\lambda <<\lambda _{C}$. When the fermion
localized in the region $\Delta x$ \ reduces its extension (increasing the
intensity of the potential, for example) then the uncertainty in the
momentum must expand, in consonance with the Heisenberg uncertainty
principle. Nevertheless, the maximum uncertainty in the momentum is
comparable with $mc$ requiring that is impossible to localize a fermion in a
region of space less than or comparable with half of its Compton wavelength
(see, for example, \cite{Greiner:1663769,strange1998relativistic}).
Nevertheless, due to the mass gain granted by its interaction with the
scalar-field background, those bound states could be highly localized
without any chance of spontaneous pair production.

\subsection{Charge conjugation}

The charge-conjugation operation is accomplished by the transformation $\psi
\rightarrow \sigma _{1}\psi ^{\ast }$ followed by $A_{0}\rightarrow -A_{0}$,
$V_{s}\rightarrow V_{s}$ and $E\rightarrow -E$ \cite{PhysRevC.73.054309}. As
a matter of fact, $A_{0}$ distinguishes fermions from antifermions but $%
V_{s} $ does not, and so the spectrum is symmetrical about $E=0$ in the case
of a pure scalar potential.

\subsection{Chiral conjugation}

The chiral-conjugation operation $\psi \rightarrow \gamma ^{5}\psi $
(according to Ref. \cite{PhysRev.106.1306}) is followed by the changes of
the signs of $V_{s}$ and $m,$ but not of $A_{0}$ and $E$ \cite%
{PhysRevC.73.054309}. One sees that the charge-conjugation and the
chiral-conjugation operations interchange the roles of the upper and lower
components of the Dirac spinor.

\subsection{Continuous chiral conjugation}

The unitary operator
\begin{equation}
U(\theta )=\exp \left( -\frac{\theta }{2}i\gamma ^{5}\right)   \label{2}
\end{equation}%
\noindent where $\theta $ is a real quantity such that $0\leq \theta \leq
\pi $, allows one to write
\begin{equation}
h\phi =E\phi   \label{22}
\end{equation}%
where

\begin{equation}
\phi =U\psi ,\quad h=UHU^{-1}  \label{23b}
\end{equation}%
with
\begin{equation}
h=\sigma _{1}cp_{1}+IA_{0}+\sigma _{3}\left( mc^{2}+V_{s}\right) \cos \theta
-\sigma _{2}\left( mc^{2}+V_{s}\right) \sin \theta  \label{333333}
\end{equation}%
It is instructive to note that the transformation preserves the form of the
current in such a way that $J^{\mu }=c\overline{\phi }\gamma ^{\mu }\phi $.
An additional important feature of the continuous chiral transformation
(see, e.g., \cite{tou1957})) induced by (\ref{2}) is that it is a symmetry
transformation when $m=V_{s}=0$.

In terms of the upper and the lower components of the spinor $\phi $,
\noindent the Dirac equation decomposes into:
\begin{equation}
\hbar c\frac{d\phi _{\pm }}{dx}\pm \left( mc^{2}+V_{s}\right) \sin \theta
\,\phi _{\pm }=i\left[ E\pm \left( mc^{2}+V_{s}\right) \cos \theta -A_{0}%
\right] \phi _{\mp }  \label{eq1}
\end{equation}
\noindent Furthermore,%
\begin{equation}
\frac{J^{0}}{c}=|\phi _{+}|^{2}+|\phi _{-}|^{2},\quad \frac{J^{1}}{c}=2\text{%
Re}\left( \phi _{+}^{\ast }\phi _{-}\right)
\end{equation}

\subsection{\noindent Special mixing}

Choosing
\begin{equation}
A_{0}=V_{s}\cos \theta  \label{5}
\end{equation}%
\noindent one has
\begin{subequations}
\begin{eqnarray}
\hbar c\frac{d\phi _{+}}{dx}+\left( mc^{2}+V_{s}\right) \sin \theta \,\phi
_{+} &=&i\left( E+mc^{2}\cos \theta \right) \phi _{-}  \label{6a} \\
&&  \notag \\
\hbar c\frac{d\phi _{-}}{dx}-\left( mc^{2}+V_{s}\right) \sin \theta \,\phi
_{-} &=&i\left[ E-\left( mc^{2}+2V_{s}\right) \cos \theta \right] \phi _{+}
\label{6b}
\end{eqnarray}%
Note that due to the constraint represented by (\ref{5}), the vector and
scalar potentials have the very same functional form and the parameter $%
\theta $ in (\ref{2}) measures the dosage of vector coupling in the
vector-scalar admixture in such a way that $|V_{s}|\geq $ $|A_{0}|$. Note
also that when the mixing angle $\theta $ goes from $\pi /2-\varepsilon $ to
$\pi /2+\varepsilon $ the sign of the spectrum undergoes an inversion under
the charge-conjugation operation whereas the spectrum of a massless fermion
is invariant under the chiral-conjugation operation. Combining
charge-conjugation and chiral-conjugation operations makes the spectrum of a
massless fermion to be symmetrical about $E=0$ in spite of the presence of
vector potential.

We now split two classes of solutions depending on whether $E$ is equal to
or different from $-mc^{2}\cos \theta $.

\subsection{The class $E\neq -mc^{2}\cos \protect\theta $}

For $E\neq -mc^{2}\cos \theta $, using the expression for $\phi _{-} $
obtained from (\ref{6a}), viz.
\end{subequations}
\begin{equation}
\phi _{-}=\frac{-i}{E+mc^{2}\cos \theta }\left[ \hbar c\frac{d\phi _{+}}{dx}%
+\left( mc^{2}+V_{s}\right) \sin \theta \,\phi _{+}\right]  \label{7}
\end{equation}
\noindent one finds%
\begin{equation}
J^{1}=\frac{2\hbar c^{2}}{E+mc^{2}\cos \theta }\,\text{Im}\left( \phi
_{+}^{\ast }\frac{d\phi _{+}}{dx}\right)
\end{equation}%
Inserting (\ref{7}) into (\ref{6b}) one arrives at the following
second-order differential equation for $\phi _{+}$:
\begin{equation}
-\frac{\hbar ^{2}}{2}\frac{d^{2}\phi _{+}}{dx^{2}}+V_{\mathtt{eff}}\,\phi
_{+}=E_{\mathtt{eff}}\,\phi _{+}  \label{8}
\end{equation}%
where%
\begin{equation}
V_{\mathtt{eff}}=\frac{\sin ^{2}\theta }{2c^{2}}V_{s}^{2}+\frac{mc^{2}+E\cos
\theta }{c^{2}}V_{s}-\frac{\hbar \sin \theta }{2c}\frac{dV_{s}}{dx}
\label{v}
\end{equation}%
and%
\begin{equation}
E_{\mathtt{eff}}=\frac{E^{2}-m^{2}c^{4}}{2c^{2}}  \label{e}
\end{equation}
\noindent Therefore, the solution of the relativistic problem for this class
is mapped into a Sturm-Liouville problem for the upper component of the
Dirac spinor. In this way one can solve the Dirac problem for determining
the possible discrete or continuous eigenvalues of the system by recurring
to the solution of a Schr\"{o}dinger-like problem because $\phi _{+}$ is a
square-integrable function.

Notice that the effective potential in (\ref{v}) behaves like $V_{s}$ in the
case of scalar and vector potentials of the same magnitude. For the case of
a pure scalar coupling ($E\neq 0$), it is also possible to write a
second-order differential equation for $\phi _{-}$ just differing from the
equation for $\phi _{+}$ in the sign of the term involving $dV_{s}/dx$,
namely,
\begin{equation}
-\frac{\hbar ^{2}}{2}\frac{d^{2}\phi _{\pm }}{dx^{2}}+\left( \frac{V_{s}^{2}%
}{2c^{2}}+mV_{s}\mp \frac{\hbar }{2c}\frac{dV_{s}}{dx}\right) \phi _{\pm
}=E_{\mathtt{eff}}\,\phi _{\pm }  \label{esc}
\end{equation}%
which is of the form of supersymmetric quantum mechanics, as has already
been appreciated in the literature \cite{Cooper19881,PhysRevA.47.1708}.

\subsection{The class $E=-mc^{2}\cos \protect\theta $}

Defining%
\begin{equation}
v\left( x\right) =\int^{x}dy\,V_{s}\left( y\right)
\end{equation}
the solutions for (\ref{6a}) and (\ref{6b}) with $E=-mc^{2}\cos \theta $ are
\begin{subequations}
\begin{eqnarray}
\phi _{+} &=&N_{+}  \label{ii1a} \\
&&  \notag \\
\phi _{-} &=&N_{-}-2\frac{i}{\hbar c}N_{+}\left[ mc^{2}x+v\left( x\right) %
\right] \cos \theta  \label{ii1b}
\end{eqnarray}%
for $\sin \theta =0$, and
\end{subequations}
\begin{subequations}
\begin{eqnarray}
\phi _{+} &=&N_{+}\exp \left\{ -\frac{\sin \theta }{\hbar c}\left[
mc^{2}x+v\left( x\right) \right] \right\}  \label{ii2a} \\
&&  \notag \\
\phi _{-} &=&N_{-}\exp \left\{ +\frac{\sin \theta }{\hbar c}\left[
mc^{2}x+v\left( x\right) \right] \right\} +i\phi _{+}\cot \theta
\label{ii2b}
\end{eqnarray}%
for $\sin \theta \neq 0$. \noindent $N_{+}$ and $N_{-}$ are normalization
constants. It is instructive to note that there is no solution for
scattering states. Both set of solutions present a space component for the
current equal to $J^{1}=2c\text{Re}\left( N_{+}^{\ast }N_{-}\right) $ and a
bound-state solution demands $N_{+}=0$ or $N_{-}=0$, because $\phi _{+}$ and
$\phi _{-}$ are square-integrable functions vanishing as $|x|\rightarrow
\infty $. It is remarkable that the eigenenergy does not depend on the
magnitude of the potential but the eigenspinor does. Note also that
\end{subequations}
\begin{equation}
\phi _{\pm }=N_{\pm }\exp \left\{ \mp \frac{1}{\hbar c}\left[
mc^{2}x+v\left( x\right) \right] \right\}  \label{sc1}
\end{equation}%
in the case of a pure scalar coupling ($E=0$) so that either $\phi _{+}=0$
or $\phi _{-}=0$.\noindent\ There is no bound-state solution for $\sin
\theta =0$, and for $\sin \theta \neq 0$ the existence of a bound state
solution depends on the asymptotic behaviour of $v(x)$ \cite%
{de2005bounded,deCastro200553}.

\section{Kink potentials}

Now we consider a kink-like potential with the asymptotic behaviour $%
V_{s}(x)\rightarrow \pm v_{0}$ as $x\rightarrow \pm x_{0}$, with $v_{0}=$
constant and $x_{0}>>\lambda $ ($\lambda $ is the range of the interaction
centered on the origin).

We turn our attention to scattering states for fermions with $E\neq
-mc^{2}\cos \theta $ coming from the left. Then, $\phi $ for $x\rightarrow
-\infty $ describes an incident wave moving to the right and a reflected
wave moving to the left, and $\phi $ for $x\rightarrow +\infty $ describes a
transmitted wave moving to the right or an evanescent wave. The upper
components for scattering states are written as
\begin{equation}
\phi _{+}=\left\{
\begin{array}{cc}
A_{+}e^{+ik_{-}x}+A_{-}e^{-ik_{-}x}, & \text{for\quad }x\rightarrow -\infty
\\
&  \\
B_{\pm }e^{\pm ik_{+}x}, & \text{for\quad }x\rightarrow +\infty
\end{array}%
\right.   \label{phi}
\end{equation}%
where
\begin{equation}
\hbar ck_{\pm }=\sqrt{\left( E\mp v_{0}\cos \theta \right) ^{2}-\left(
mc^{2}\pm v_{0}\right) ^{2}}
\end{equation}%
Note that $k_{+}$ is a real number for a progressive wave and an imaginary
number for an evanescent wave ($k_{-}$ is a real number for scattering
states). Therefore,%
\begin{equation}
J_{\gtrless }^{1}\left( -\infty \right) =\frac{2\hbar c^{2}k_{-}}{%
E+mc^{2}\cos \theta }\left( |A_{\pm }|^{2}-|A_{\mp }|^{2}\right) ,\text{%
\quad for\quad }E\gtrless -mc^{2}\cos \theta
\end{equation}%
and
\begin{equation}
J_{\gtrless }^{1}\left( +\infty \right) =\pm \,\frac{2\hbar c^{2}\text{Re}%
\,k_{+}}{E+mc^{2}\cos \theta }\,|B_{\pm }|^{2},\text{\quad for\quad }%
E\gtrless -mc^{2}\cos \theta
\end{equation}%
Note also that%
\begin{equation}
J_{\gtrless }^{1}\left( -\infty \right) =J_{\mathtt{inc}}-J_{\mathtt{ref}}
\end{equation}%
and%
\begin{equation}
J_{\gtrless }^{1}\left( +\infty \right) =J_{\mathtt{tran}}
\end{equation}%
where $J_{\mathtt{inc}}$, $J_{\mathtt{ref}}$ and $J_{\mathtt{tran}}$ are
nonnegative quantities characterizing the incident, reflected and
transmitted waves, respectively. Note also that the roles of $A_{+}$ and $%
A_{-}$ are exchanged as the sign of $E+mc^{2}\cos \theta $ changes. In fact,
if $E>-mc^{2}\cos \theta $, then $A_{+}e^{+ik_{-}x}$ ($A_{-}e^{-ik_{-}x}$)
will describe the incident (reflected) wave, and $B_{-}=0$. On the other
hand, if $E<-mc^{2}\cos \theta $, then $A_{-}e^{-ik_{-}x}$ ($%
A_{+}e^{+ik_{-}x}$) will describe the incident (reflected) wave, and $B_{+}=0
$. Therefore, the reflection and transmission amplitudes are given by%
\begin{equation}
r_{\gtrless }=\frac{A_{\mp }}{A_{\pm }},\text{\quad }t_{\gtrless }=\frac{%
B_{\pm }}{A_{\pm }},\text{\quad for\quad }E\gtrless -mc^{2}\cos \theta
\label{t}
\end{equation}%
To determine the transmission coefficient we use the current densities $%
J_{\gtrless }^{1}\left( -\infty \right) $ and $J_{\gtrless }^{1}\left(
+\infty \right) $. The $x$-independent space component of the current allows
us to define the reflection and transmission coefficients as%
\begin{equation}
R_{\gtrless }=\frac{|A_{\mp }|^{2}}{|A_{\pm }|^{2}},\text{\quad }T_{\gtrless
}=\frac{\text{Re}\,k_{+}}{k_{-}}\frac{|B_{\pm }|^{2}}{|A_{\pm }|^{2}},\text{%
\quad for\quad }E\gtrless -mc^{2}\cos \theta   \label{tr}
\end{equation}%
Notice that $R_{\gtrless }+T_{\gtrless }=1$ by construction. The possibility
of bound states requires a solution with an asymptotic behaviour given by (%
\ref{phi}) with $k_{\pm }=i|k_{\pm }|$ and $A_{+}=B_{-}=0$, or $k_{\pm
}=-i|k_{\pm }|$ and $A_{-}=B_{+}=0$, to obtain a square-integrable $\phi _{+}
$. On the other hand, if one considers the transmission amplitude $t$ in (%
\ref{t}) as a function of the complex variables $k_{\pm }$ one sees that for
$k_{\pm }>0$ one obtains the scattering states whereas the bound states
would be obtained by the poles lying along the imaginary axis of the complex
$k$-plane.

As for $E=-mc^{2}\cos \theta $, the existence of a bound-state solution
requires $|v_{0}|>mc^{2}$ so that the eigenspinor behaves asymptotically as%
\begin{equation}
\phi \sim \left(
\begin{array}{c}
1 \\
i\cot \theta%
\end{array}%
\right) f
\end{equation}%
for $v_{0}>mc^{2}$, and%
\begin{equation}
\phi \sim \left(
\begin{array}{c}
0 \\
1%
\end{array}%
\right) f
\end{equation}%
for $v_{0}<-mc^{2}$. Here,
\begin{equation}
f=\exp \left\{ -\frac{\sin \theta }{\hbar c}\left[ |v_{0}|+mc^{2}\text{sgn}%
\left( v_{0}x\right) \right] |x|\right\}  \label{efe}
\end{equation}

Armed with the knowledge about asymptotic solutions and with the definition
of the transmission coefficient we should proceed for searching solutions on
the entire region of space. Nevertheless, we can not tell much more about
the problem until the potential function is specified.

\section{The smooth step potential}

Now the scalar potential takes the form%
\begin{equation}
V_{s}=v_{0}\,\mathrm{\tanh }\,x\!/\!\lambda  \label{vs}
\end{equation}%
where the skew positive parameter $\lambda $ is related to the range of the
interaction which makes $V_{s}$ to change noticeably in the interval $%
-\lambda <x<\lambda $, and $v_{0}$ is the height of the potential at $%
x=+\infty $. Notice that as $\lambda \rightarrow 0$, the case of an extreme
relativistic regime, the smooth step approximates the sign potential already
considered in Ref. \cite{Castilho20141}.

As commented before, there is no isolated solution from the Sturm-Liouville
problem when $\sin \theta =0$, and the existence of a well-behaved isolated
solution when $\sin \theta \neq 0$ makes%
\begin{equation}
v\left( x\right) =\lambda v_{0}\ln \left( \cosh x\!/\!\lambda \right)
\label{vx}
\end{equation}
and requires $|v_{0}|>mc^{2}$:%
\begin{equation}
\phi =\left(
\begin{array}{c}
1 \\
i\cot \theta%
\end{array}%
\right) N_{>}\,f  \label{s1}
\end{equation}%
for $v_{0}>mc^{2}$, and%
\begin{equation}
\phi =\left(
\begin{array}{c}
0 \\
1%
\end{array}%
\right) N_{<}\,f  \label{s2}
\end{equation}%
for $v_{0}<-mc^{2}$. Here,
\begin{equation}
f=\frac{\exp \left( -\alpha _{1}x\right) }{\cosh ^{\alpha _{2}}x\!/\!\lambda
}
\end{equation}%
where%
\begin{equation}
\alpha _{1}=\frac{\text{sgn}\left( v_{0}\right) mc\sin \theta }{\hbar }%
,\quad \alpha _{2}=\frac{\lambda |v_{0}|\sin \theta }{\hbar c}
\end{equation}%
The normalization condition%
\begin{equation}
\int_{-\infty }^{+\infty }dx\,\left( |\phi _{+}|^{2}+|\phi _{-}|^{2}\right)
=1  \label{norma}
\end{equation}%
and the integral tabulated (see the formula 3.512.1, or 8.380.10, in Ref.
\cite{GradshteynRyzhik198006})%
\begin{equation}
\int_{0}^{\infty }dx\,\frac{\cosh 2\beta _{1}x}{\cosh ^{2\beta _{2}}\gamma x}%
=\frac{2^{2\beta _{2}}}{4\gamma }B\left( \beta _{2}+\frac{\beta _{1}}{\gamma
},\beta _{2}-\frac{\beta _{1}}{\gamma }\right)
\end{equation}%
where $B\left( z_{1},z_{2}\right) $ is the beta function \cite%
{AbramowitzStegun196506}, allow one to determine $N_{\gtrless }$. In the way
indicated one can find the position probability density \cite%
{Castilho2014164}
\begin{equation}
|\phi |^{2}=\frac{2f^{2}}{2^{2\alpha _{2}}\lambda B\left( \alpha _{+},\alpha
_{-}\right) }
\end{equation}%
where%
\begin{equation}
\alpha _{\pm }=\alpha _{2}\pm \lambda \alpha _{1}
\end{equation}%
Therefore, a massive fermion tends to concentrate at the left (right) region
when $v_{0}>0$ ($v_{0}<0$), and tends to avoid the origin more and more as $%
\sin \theta $ decreases. A massless fermion has a position probability
density symmetric around the origin. One can see that the best localization
occurs for a pure scalar coupling. In fact, the fermion becomes delocalized
as $\sin \theta $ decreases.

The expectation value of $x$ and $x^{2}$ is given by%
\begin{equation}
<x>=-\frac{4}{2^{2\alpha _{2}}\lambda B\left( \alpha _{+},\alpha _{-}\right)
}\int_{0}^{\infty }dx\,\frac{x\sinh 2\alpha _{1}x}{\cosh ^{2\alpha
_{2}}x\!/\!\lambda }  \label{espec1}
\end{equation}%
and%
\begin{equation}
<x^{2}>=\frac{4}{2^{2\alpha _{2}}\lambda B\left( \alpha _{+},\alpha
_{-}\right) }\int_{0}^{\infty }dx\,\frac{x^{2}\cosh 2\alpha _{1}x}{\cosh
^{2\alpha _{2}}x\!/\!\lambda }  \label{espec2}
\end{equation}%
Defining
\begin{subequations}
\begin{eqnarray}
{\Delta }\left( \alpha \right) &=&\boldsymbol{\psi \,}\left( \alpha
_{+}\right) -\boldsymbol{\psi \,}\left( \alpha _{-}\right)  \label{44a} \\
&&  \notag \\
{\Sigma }^{\left( 1\right) }\left( \alpha \right) &=&\boldsymbol{\psi \,}%
^{\left( 1\right) }\left( \alpha _{+}\right) +\boldsymbol{\psi \,}^{\left(
1\right) }\left( \alpha _{-}\right)  \label{44b}
\end{eqnarray}%
where
\end{subequations}
\begin{equation}
\boldsymbol{\psi}\left( z\right) =\frac{d\ln \Gamma \left( z\right) }{dz}
\label{di}
\end{equation}%
is the digamma (psi) function and%
\begin{equation}
\boldsymbol{\psi}^{\left( 1\right) }\left( z\right) =\frac{d\boldsymbol{\psi}%
\left( z\right) }{dz}  \label{tri}
\end{equation}%
is the trigamma function \cite{AbramowitzStegun196506}, printed in a
boldface type to differ from the Dirac eigenspinor, and using a pair of
integrals tabulated in \cite{Castilho2014164}, $<x>$ and $<x^{2}>$ can be
simplified to
\begin{equation}
<x>=-\frac{\lambda }{2}{\Delta }\left( \alpha \right)
\end{equation}%
and%
\begin{equation}
<x^{2}>=\frac{\lambda ^{2}}{4}\,{\Sigma }^{\left( 1\right) }\left( \alpha
\right) +<x>^{2}
\end{equation}%
and hence the fermion is confined within an interval%
\begin{equation}
\Delta x=\sqrt{<x^{2}>-<x>^{2}}  \label{ddx}
\end{equation}%
given by%
\begin{equation}
\Delta x=\frac{\lambda }{2}\sqrt{{\Sigma }^{\left( 1\right) }\left( \alpha
\right) }  \label{dx}
\end{equation}%
With the help of a few approximate formulas for the special functions in
Ref. \cite{Castilho2014164}, one obtains the values for $<x>$ and $\Delta x$
either in the case of $\sin \theta \simeq 0$ or $\lambda <<\lambda _{C}$:
\begin{subequations}
\begin{eqnarray}
<x> &\simeq &-\text{sgn}\left( v_{0}\right) \frac{\hbar c}{\sin \theta }%
\frac{mc^{2}}{v_{0}^{2}-m^{2}c^{4}}  \label{g1a} \\
&&  \notag \\
\Delta x &\simeq &\frac{\hbar c}{\sqrt{2}\sin \theta }\frac{\sqrt{%
v_{0}^{2}+m^{2}c^{4}}}{v_{0}^{2}-m^{2}c^{4}}  \label{g1b}
\end{eqnarray}%
One can also see that when $\lambda >>\lambda _{C}$ or $|v_{0}|>>mc^{2}$
\end{subequations}
\begin{subequations}
\begin{eqnarray}
<x> &\simeq &\frac{\lambda }{2}\ln \frac{|v_{0}|-\text{sgn}\left(
v_{0}\right) mc^{2}}{|v_{0}|+\text{sgn}\left( v_{0}\right) mc^{2}}
\label{g2a} \\
&&  \notag \\
\Delta x &\simeq &\sqrt{\frac{\lambda \hbar c}{2\sin \theta }\frac{|v_{0}|}{%
v_{0}^{2}-m^{2}c^{4}}}  \label{g2b}
\end{eqnarray}%
Again one can see that the fermion becomes delocalized as $\sin \theta $
decreases and that the best localization occurs for a pure scalar coupling.
More than this, $<x>\rightarrow -\infty $ and $\Delta x\rightarrow \infty $
as $|v_{0}|\rightarrow mc^{2}$, and besides $<x>\rightarrow 0$ and $\Delta
x\rightarrow 0$ as $|v_{0}|\rightarrow \infty $.

Those last results show that $\Delta x$ reduces its extension with rising $%
|v_{0}|$ or $\sin \theta $, or decreasing $\lambda $, in such a way that $%
\Delta x$ can be arbitrarily small for a potential strong enough or
short-range enough. The impasse related to the Heisenberg uncertainty
principle can be broken by resorting to the concepts of effective mass and
effective Compton wavelength. Indeed, if one defines an effective mass as
\end{subequations}
\begin{equation}
m_{\mathtt{eff}}=m\sqrt{1+\left( \frac{v_{0}}{mc^{2}}\right) ^{2}}
\label{mef}
\end{equation}%
and an effective Compton wavelength $\lambda _{\mathtt{eff}}=\hbar /\left(
m_{\mathtt{eff}}c\right) $, one will find
\begin{equation}
\Delta x=\frac{\sqrt{2}\lambda _{\mathtt{eff}}}{4\sin \theta }\sqrt{\left(
\alpha _{+}^{2}+\alpha _{-}^{2}\right) {\Sigma }^{\left( 1\right) }\left(
\alpha \right) }  \label{dx1}
\end{equation}%
It follows that the high localization of fermions, related to high values of
$|v_{0}|$ or small values of $\lambda $, never menaces the single-particle
interpretation of the Dirac theory even if the fermion is massless ($m_{%
\mathtt{eff}}=|v_{0}|/c^{2}$). As a matter of fact, (\ref{g1b}) furnishes%
\begin{equation}
\left( \Delta x\right) _{\min }\simeq \frac{\lambda _{\mathtt{eff}}}{\sqrt{2}%
\sin \theta }  \label{dd}
\end{equation}%
for $|v_{0}|>>mc^{2}$ or $\lambda <<\lambda _{C}$.

\section{Final remarks}

After reviewing the use of a continuous chiral transformation for solving
the Dirac equation in the background of scalar and vector potential
developed in Refs. \cite{Castilho20141}-\cite{Castilho2014164}, we have done
an extension for arbitrary kink-like potentials which generalizes the
Jackiw-Rebbi \cite{PhysRevD.13.3398} model not only for considering massive
fermions but also for taking into account an additional vector coupling.
Concentrating our attention on the \textquotedblleft
zero-mode\textquotedblright\ solution, we have shown that, due to the
sizeable additional mass acquired by the fermion resulting from its
interaction with the scalar-field background, those bound states can be
highly localized by very strong potentials without any chance of spontaneous
pair production. This fact is convincing because the scalar is stronger than
the vector coupling, and so the conditions for Klein's paradox are never
reached.

\ack This work was supported in part by means of funds provided by CNPq.

\section*{References}

\bibliographystyle{plain}
\bibliography{newbib}

\end{document}